\documentstyle[12pt]{article}
\def\be {\begin{equation}}
\def\ee {\end{equation}}
\def\ba {\begin{eqnarray}}
\def\ea {\end{eqnarray}}
\newcommand{\bq}{\begin{eqnarray}}
\newcommand{\eq}{\end{eqnarray}}

\def\bi {\begin{itemize}}
\def\ei {\end{itemize}}
\begin{document}
\def\bea{\begin{eqnarray}}
\def\eea{\end{eqnarray}}
\title{\bf {Reply to the comments of Karami on the Interacting holographic dark energy
model and generalized second law of thermodynamics in a non-flat
universe}}
 \author{M.R. Setare  \footnote{E-mail: rezakord@ipm.ir}
  \\ {Department of Physics, University of Kurdistan, Pasdaran Ave., Sanandaj, Iran}}
\date{\small{}} \maketitle
\begin{abstract}
In \cite{2} K. Karami commented on my publication \cite{1} and
claimed that my result would be invalid. In fact he found an
error in calculation of the entropy of cold dark matter in
section 3 of my article, which is, however, a rather trivial one.
It is shown that my discussion in section 3 of my paper works
correctly if we take the value of free parameter $b^2>0.264$.
 \end{abstract}

\newpage
\section{Reply}
The comment of author \cite{2} is based on an error in the sign of
first term in Eq.(35) of paper \cite{1}. Due to this error the
result of equations (37) and (38) also will change. The correct
form is given by Eq.(4) in \cite{2}. But this correction do not
change my result in paper \cite{1}. In the paper \cite{1} I have
claimed that the generalized second law (GSL) of thermodynamics
can be satisfy in non-flat universe enclosed by the event horizon
measured from the sphere of the event horizon $L$. In my formula
(41), or in Eq.(6) of \cite{2} which is correct version of (41)
due the correction on eq.(35), there are some parameters,
$\Omega_k$, $\Omega_\Lambda$, $\cos y$, $c$, and $b^2$. \\
The holographic dark energy model has been tested and constrained
by various astronomical observations, in both flat and non-flat
cases. These observational data include type Ia supernovae,
cosmic microwave background, baryon acoustic oscillation, and the
X-ray gas mass fraction of galaxy clusters. According to the
analysis of the observational data for the holographic dark
energy model, we find that generally $c<1$. Here we summarized
the main constraint results as follows:

\begin{enumerate}

\item For flat universe, using only the SN data to constrain the holographic
dark energy model, we get the fit results:
$c=0.21^{+0.41}_{-0.12}$, $\Omega_{\rm m0}=0.47^{+0.06}_{-0.15}$,
with the minimal chi-square corresponding to the best fit
$\chi^2_{\rm min}=173.44$ \cite{obs1}. When combining the
information from SN Ia \cite{Riess:2004nr}, CMB \cite{CMB} and
BAO \cite{BAO}, the fitting for the holographic dark energy model
gives the parameter constraints in 1 $\sigma$:
$c=0.81^{+0.23}_{-0.16}$, $\Omega_{\rm m0}=0.28\pm 0.03$, with
$\chi_{\rm min}^2=176.67$ \cite{obs1}.

\item Also for the flat case, the X-ray gas mass
fraction of rich clusters, as a function of redshift, has also
been used to constrain the holographic dark energy model
\cite{obs2}. The main results, i.e. the 1 $\sigma$ fit values for
$c$ and $\Omega_{\rm m0}$ are: $c=0.61^{+0.45}_{-0.21}$ and
$\Omega_{\rm m0}=0.24^{+0.06}_{-0.05}$, with the best-fit
chi-square $\chi_{\rm min}^2=25.00$ \cite{obs2}.

\item For the non-flat universe, the authors of \cite{obsnonflat} used the data
coming from the SN and CMB to constrain the holographic dark
energy model, and got the 1 $\sigma$ fit results:
$c=0.84^{+0.16}_{-0.03}$, $\Omega_{m0}=0.29^{+0.06}_{0.08}$, and
$\Omega_{k0}=0.02\pm 0.10$, with the best-fit chi-square
$\chi_{\rm min}^2=176.12$.

\end{enumerate}
Due to these result, and such as \cite{1}  we can assume the value
of $\Omega_k$, $\Omega_\Lambda$, $\cos y$, and $c$ in present time
 respectively as, $0.01$, $0.73$, $0.99$, and $1$. In the other
 hand $b$ is a dimensionless coupling constant, which is a free
 parameter, where appear in Eq.(21) in \cite{1}. If we substitute
 the mentioned values for $\Omega_k$, $\Omega_\Lambda$, $\cos y$,
 $c=1$
 in Eq.(6) of \cite{2} we obtain following relation
 \begin{equation}
\frac{d}{dx}(S_{\Lambda}+
S_m+S_L)=\frac{\pi^2M_{p}^{2}}{H^{2}}(3.38-2.86-10.732+38.63 b^2)
 \end{equation}
 easily one can see if we take the value of $b^2>0.264$, then we
 obtain $\frac{d}{dx}(S_{\Lambda}+
S_m+S_L)>0$. It should be stressed that evidence was recently
provided by the Abell Cluster A586 in support of the interaction
between dark energy and dark matter \cite{ab}. However, despite
the fact that numerous works have been performed till now, there
are no strong observational bounds on the strength of this
interaction \cite{ab1}. This weakness to set stringent
(observational or theoretical) constraints on the strength of the
coupling between dark energy and dark matter stems from our
unawareness of the nature and origin of dark components of the
Universe. It is therefore more than obvious that further work is
needed to this direction. In any case may be we can said that for
the validity of GSL in the framework of our model, the coupling
constant between dark energy and cold dark
matter must satisfy $b^2>0.264$ in present time. \\
 So in contrast to the claim of author \cite{2}, the
essential conclusion of \cite{1} is true and the GSL can be valid
for the present time for the interacting holographic dark energy
with cold dark matter in a non-flat universe enveloped by the
event horizon measured from the sphere of the horizon named $L$.\\
Finally it should be mentioned that recently we have investigated
the validity of the generalized second law of thermodynamics, in
the cosmological scenario where dark energy interacts with both
dark matter and radiation \cite{jss}. Then we have shown that the
generalized second law is always and generally valid, as long as
one considers the apparent horizon as the universe ``radius''
(the use of other choices, such is the future event horizon,
leads to conditional validity only), independently of the specific
interaction form, of the fluids equation-of-state parameters and
of  the background geometry.


\begin{thebibliography}{99}
\bibitem{1}M. R. Setare, J. Cosmol. Astropart. Phys. 01, 023 (2007).
\bibitem{2}K. Karami, arXiv:0911.4808 [physics.gen-ph].
\bibitem{obs1}
  X.~Zhang and F.~Q.~Wu,
  Phys.\ Rev.\ D {\bf 72}, 043524 (2005)
  [astro-ph/0506310].
\bibitem{CMB}
  D.~N.~Spergel {\it et al.}  [WMAP Collaboration],
  Astrophys.\ J.\ Suppl.\  {\bf 148}, 175 (2003)
  [astro-ph/0302209];\\
  D.~N.~Spergel {\it et al.},
  astro-ph/0603449.
\bibitem{Riess:2004nr}
  A.~G.~Riess {\it et al.}  [Supernova Search Team Collaboration],
  Astrophys.\ J.\  {\bf 607}, 665 (2004)
  [astro-ph/0402512].



\bibitem{BAO}
  D.~J.~Eisenstein {\it et al.}  [SDSS Collaboration],
  Astrophys.\ J.\  {\bf 633}, 560 (2005)
  [astro-ph/0501171].
\bibitem{obs2}
  Z.~Chang, F.~Q.~Wu and X.~Zhang,
  Phys.\ Lett.\ B {\bf 633}, 14 (2006)
  [astro-ph/0509531].
  \bibitem{obsnonflat}
  Y.~G.~Gong, B.~Wang and Y.~Z.~Zhang,
  Phys.\ Rev.\ D {\bf 72}, 043510 (2005)
  [hep-th/0412218].
  \bibitem{ab}O. Bertolami, F. Gil Pedro and M. Le Delliou, Phys. Lett. B 654, 165 (2007)
[arXiv:astro-ph/0703462]; M. Le Delliou, O. Bertolami and F. Gil
Pedro, AIP Conf. Proc. 957, 421 (2007) [arXiv:0709.2505
[astro-ph]].
  \bibitem{ab1}L. Amendola, G. Camargo Campos and R. Rosenfeld, Phys. Rev. D 75, 083506
(2007) [arXiv:astro-ph/0610806]; Z. K. Guo, N. Ohta and S.
Tsujikawa, Phys. Rev. D 76, 023508 (2007) [arXiv:astroph/
0702015]; C. Feng, B. Wang, Y. Gong and R. K. Su, JCAP 0709, 005
(2007) [arXiv:0706.4033 [astro-ph]]; E. Abdalla, L. R. W. Abramo,
L. . J. Sodre and B. Wang, arXiv:0710.1198 [astro-ph].
\bibitem{jss}M. Jamil, E. N. Saridakis, M. R. Setare, arXiv:0910.0822v1
[hep-th], accepted for publication in Phys. Rev. D (2010).
\end{thebibliography}
\end{document}